\documentclass[12pt,floatfix]{article}
\usepackage{mathrsfs}

\usepackage{graphicx}
\usepackage{bm}

\begin{document}

\title{Property of the low-lying states at the critical point
of the phase transition in U(4) vibron model}



\author{{Ze-bo Li$^{a}$, Xia-ping Tang$^{a}$, Yu Zhang$^{a}$, Xing-chen Yang$^{a}$, Zhao
Wang$^{a}$,}\\
{and Yu-xin Liu$^{a,b,}$\thanks{Corresponding author, e-mail address: yxliu@pku.edu.cn} }\\[3mm]
\normalsize{$^a$ Department of Physics and State Key Laboratory of
Nuclear Physics and Technology,}\\
\normalsize{ Peking University, Beijing 100871, China}\\
\normalsize{$^c$ Center of Theoretical Nuclear Physics, National
Laboratory of Heavy Ion Accelerator,}\\ \normalsize{ Lanzhou 730000,
China}  }

%
%
%
%
%
%
%

\maketitle

\begin{abstract}
We study the properties of the low-lying states at the critical
point of the phase transition from U(3) to O(4) symmetry in the U(4)
vibron model in detail. By analyzing the general characteristics and
comparing the calculated results of the energy spectra and the E1,
E2 transition rates in E(3) symmetry, in $r^{4}$ potential model and
the finite boson number case in boson space, we find that the
results in the $r^{4}$ potential demonstrates the characteristic of
the classical limit at the critical point well and the E(3) symmetry
over-predict the energy levels and under-predict the E1 and E2
transition rates of the states at the critical point. However, the
E(3) symmetry may describe part of the properties of the system with
boson number around 10 to 20. We also confirm that the
${^{12}}$C+${^{12}}$C system is an empirical evidence of the state
at the critical point of the phase transition in the U(4) model when
concerning the energies of the low-lying resonant states.

\end{abstract}


%
%



\newpage

\parindent=20pt


Quantum phase transitions in mesoscopic system (system with a finite
number of particles $N$), such as atomic
nuclei~\cite{GK1980,DSI1980,FGD1981,IC1981,WC1983,IachelloBook87,Leviatan,Warner02,Jolie2002,LG03,Regan03,Panetc035,Iachello2004,Rowe2004,Heenen05,Cejnar2005,Dusuel2005,Casten067,Ring07,Liu2006,Arias2007,Liu2007,Bonatsos08},
molecules~\cite{IachelloBook95,Kuyucak1999,Iachello2005}, atomic
clusters~\cite{Hess2006} and finite polymers have recently been
attracting a lot of interests. The transition in these systems are
among different shapes, geometric configurations, and modes of
collective motions.
For nuclei, it has been well known that there exists vibrational,
$\gamma$-soft rotational, axially rotational, and other collective
modes.
The interacting bosons model (IBM, the simplest one is the U(6)
model including $s$- and $d$-bosons)~\cite{IachelloBook87} has been
shown to be successful in studying the properties of the low-lying
collective states of even-even nuclei and the shape phase
transitions
(see for example,
Refs.~\cite{GK1980,DSI1980,FGD1981,IC1981,WC1983,IachelloBook87,Leviatan,Warner02,Jolie2002,LG03,Panetc035,Iachello2004,Rowe2004,Cejnar2005,Dusuel2005,Casten067,Liu2006,Arias2007,Liu2007,Bonatsos08}
).
It has also been well known that there exhibits vibration and
rotation in molecules.
To characterize the relative motion of a dipole-deformation in the
three dimensional space and describe the behavior of rotational and
vibrational motions of molecules, the U(4) vibron model has been
developed~\cite{IachelloBook95,Iachello1982} and applied to two-body
(or two-cluster) systems such as diatomic
molecules~\cite{Iachello1982,Kuyucak958}, binary
clusters~\cite{Iachello1981,Erb1981,Cseh1985,Cseh1993}, $q\bar{q}$
mesons~\cite{Iachello1991,Pan2006}, and so on. It has been shown
that the U(4) model involves two dynamical symmetries, namely U(3)
and O(4), and there exists a second order phase transition from U(3)
to O(4) symmetry~\cite{Dieperink1982,Liu2008}.
For the properties of the states at the critical point of the
vibrational to $\gamma$-soft rotational, the vibrational to axially
rotational phase transition of nucleus, respectively, there have had
thorough studies in theory (see, for example,
Refs.~\cite{LG03,Iachello2000,Iachello2001,Arias2003,Bonatsos2004,Ramos2005,McCutchan2006,Arias2008})
and quite a lot of empirical evidences have also been found (see,
for instance, Refs.~\cite{Casten00,Casten01}). For the properties of
the states around the critical point of the U(3) and O(4) phase
transition in the U(4) model, even though they can be described
approximately with the E(3) symmetry, which has been discussed in
some sense~\cite{Liu2008}, they have not yet been investigated in
detail. We will then study the low-lying energy spectrum and the E1,
E2 transition rates of the low-lying states at the critical point of
the phase transition in the U(4) vibration model in this paper.
%

%
%
In the U(4) model, elementary excitations are dipole p-bosons with
spin and parity $J^{\pi}=1^{-}$ and scalar s-bosons with
$J^{\pi}=0^{+}$. With assumptions that the total number of bosons
and the angular momentum of the system are conserved, there are only
two dynamical symmetry limits, U(3) and O(4). Accordingly, there
exist two dynamical symmetry chains:
\begin{eqnarray}
  U(4)\supset U(3)\supset O(3), & ~~~~~~~~~~~({\rm I})  & \\
  U(4)\supset O(4)\supset O(3). & ~~~~~~~~~~~({\rm II}) &
\end{eqnarray}
It has been shown that the U(3) symmetry corresponds to nonrigid
ro-vibrations, while the O(4) symmetry represents rigid
ro-vibrations~\cite{Iachello1982}. A general Hamiltonian of the U(4)
vibron model with only one- and two-body interactions being taken
into account can be expressed in terms of the linear and quadratic
invariant operators (Casimir operators) of all the subgroups
contained in the dynamical group chains.

To study the property of the phase transition between U(3) symmetry
and O(4) symmetry, one starts usually from the simple Hamiltonian
\begin{equation} \label{Hamiltonian}
\hat{H} = \varepsilon[(1- \eta )\hat{n}-{ \frac{\eta}{f(N)}
}\hat{D}\cdot\hat{D}] \, ,
\end{equation}
where $\varepsilon$ is a scale parameter and it can be taken as one
for convenience without any loss of generality. $\hat{n}=\sum_{m}
p^{\dag}_{m} p_{m}$ is the number operator of p-bosons,
$\hat{D}^{(1)}_q=(s^{\dag} \tilde{p} + p^{\dag} \tilde{s})_q^{(1)}$
is the electric dipole operator, where $\tilde{s}=s$ and
$\tilde{p}_m=(-1)^{1-m}p_{-m}$. $f(N)$ is a linear function of total
boson number $N$. $\eta$ is the control parameter. It is easy to
show that such a Hamiltonian can be alternatively written as

\begin{equation} \label{Hamiltonian-Casimir}
\hat{H} = \varepsilon(1- \eta )C_{1U(3)} - \varepsilon {
\frac{\eta}{f(N)} } C_{2O(4)} + \varepsilon { \frac{\eta}{f(N)} }
C_{2O(3)} \, .
\end{equation}
It is obvious that the system is in U(3) symmetry when $\eta=0$, and
in O(4) symmetry if $\eta=1$. By varying $\eta\in [0,1]$, we can
realize the U(3)-O(4) phase transition.

The classical limit corresponding to the Hamiltonian in
Eq.~(\ref{Hamiltonian}) can be obtained by considering its
expectation value of the coherent state~\cite{IachelloBook95}
\begin{equation} \label{Coherent-state}
|N;\mathbf{t}\rangle=(N!)^{-1/2}[(1-\mathbf{t}^{*}\cdot
\mathbf{t})^{1/2}s^{\dag}+{\mathbf{t}} \cdot
\mathbf{p}^{\dag}]^N|0\rangle \, ,
\end{equation}
where $\mathbf{t}$ is a complex three-dimensional vector, and its
complex conjugate is denoted by $\mathbf{t}^*$. Then the classical
Hamiltonian can be given as $H_{cl}=\langle N;\mathbf{ti}\mid
\hat{H} \mid N;\mathbf{t}\rangle$. One can introduce canonical
position and momentum variables $\mathbf{r}$ and $\mathbf{q}$ (they
can be the ones in three dimension) by the
transformation~\cite{IachelloBook95}

\begin{equation}  \label{collective-varaibles}
\mathbf{t}=(\mathbf{r} + i\mathbf{q} )/\sqrt{2}\, , ~~~~~~
\mathbf{t}^{*}=( \mathbf{r} - i\mathbf{q} )/\sqrt{2}\, .
\end{equation}
So the classical potential is just the value of $H_{cl}(q,r)$ with
$q=0$, where $|\mathbf{r}|=r$ and $|\mathbf{q}|=q$ i.e.,
\begin{equation}
V(r)=H_{cl}(q=0,r)\, .
\end{equation}
In the case of taking $f(N)$ in Eq.~(\ref{Hamiltonian}) to be $3N$,
the energy potential corresponding to the Hamiltonian in
Eq.~(\ref{Hamiltonian}) can be concretely expressed as
\begin{equation} \label{Potential}
V(r)=N \Big[(1- \eta ){\frac{r^2}{2}}-{ \frac{\eta}{3}} r^2(2-r^2)
\Big] \, .
\end{equation}

From Eq.~(\ref{Potential}), we could easily recognize that the
classical potential for U(3) symmetry (with $\eta=0$) is
$V_{U(3)}(r)= \frac{N}{2} r^2$, while the classical potential for
O(4) symmetry (with $\eta=1$) is $V_{O(4)}(r)=-\frac{N}{3}
r^2(2-r^2)$. By analyzing the stability of
 the system with potential in Eq.~(\ref{Potential}) we
could find that the critical point of the quantum transition
corresponds to the control parameter $\eta_c=\frac{3}{7}$. It is
obvious that, as $\eta < \frac{3}{7} $, $V_{min}(r)=0$; if $\eta
>\frac{3}{7}$, $V_{min}(r)=-\frac{N}{48} \frac{(7 \eta -3)^{2}}{\eta} $.
Furthermore, at the critical point with control parameter
$\eta=\eta_{c} = \frac{3}{7} $, the energy potential can be
explicitly written as
\begin{equation}\label{Potential-criticalpoint}
V_{cri}(r)= \frac{N}{7} r^4 \, ,
\end{equation}
and the Hamiltonian corresponding to Eq.~(\ref{Hamiltonian}) could
be simplified to a Schr\"{o}dinger equation
\begin{equation}\label{Schrodinger}
\hat{H}\Phi= \Big[ \frac{\hat{p}^{2}}{2m} + \frac{r^{4}}{7} \Big]
\Phi = E \Phi \, .
\end{equation}

Looking through the characteristic of the potential at the critical
point in Eq.~(\ref{Potential-criticalpoint}), one can learn that it
is quite flat around its bottom. To study the property of the states
at the critical point of the phase transition in the U(4) model, at
first one can go along the way taken in Ref.~\cite{Iachello2000} and
approximate the potential around the critical point to be a
three-dimensional infinite square well
\begin{equation}\label{well}
V(r) =  \left\{ \begin{array}{cc} 0 \, , & ~~~~ r \leq r_{W} \, ,  \\
\infty \, , & ~~~~ r > r_{W} \, . \end{array} \right.
\end{equation}
It is apparent that such a potential deviates from that of the E(5)
symmetry of the transition from U(5) to O(6) symmetry in the
IBM~\cite{Iachello2000} only in dimension. Then the states generated
from this potential can de denoted as the ones with the E(3)
symmetry.

\begin{figure}[h!]
\begin{center}
\includegraphics[scale=1.0]{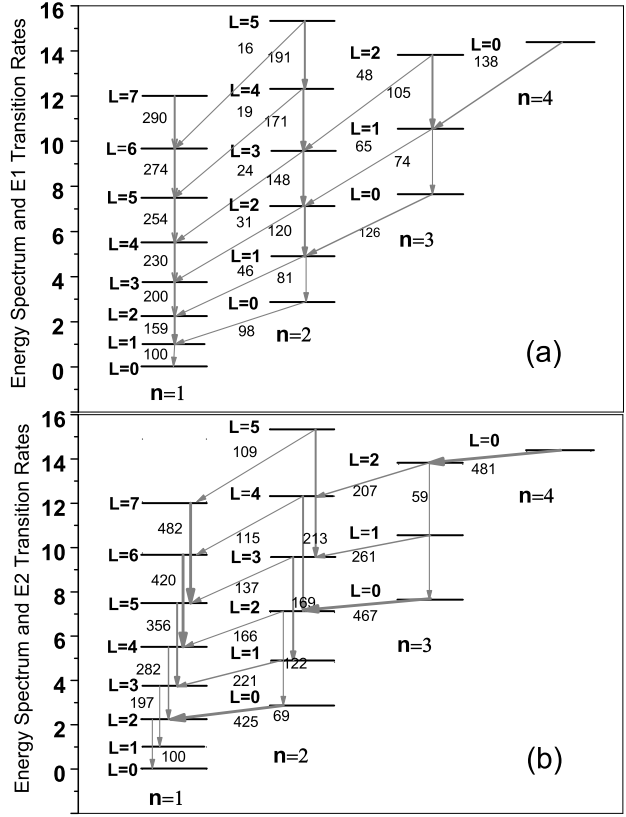}
\vspace*{-5mm}\caption{Calculated energy spectrum of the low-lying
states in E(3) symmetry (with a potential of infinite square well)
around the critical point of the phase transition U(3)-O(4) in the
U(4) vibron model and the corresponding E1 (panel (a)), E2 (panel
(b)) transition rates (the numbers close to the arrow are the value
of the rate with normalization  $B(\mbox{E1}; 1^{-}_{1} \rightarrow
0^{+}_{1}) = 100 $, $B(\mbox{E2}; 2^{+}_{1} \rightarrow 0^{+}_{1} )
= 100$, respectively). } \label{F1}
\end{center}
\end{figure}

In such a situation, the Schr\"{o}dinger equation is exactly
solvable and the solution could be expressed as some Bessel
function. The excitation energy can be given as
\begin{equation}
E_{n,L}=\frac{2B}{\hbar}k^2_{n,L} \, ,
\end{equation}
with $k^2_{n,L}=\frac{y_{n,L}}{x_W}$, where $y_{n,L}$ is the $n$th
zero point of the Bessel function $J_{L+1/2}(z)$, $B$ is a constant.
We give in Fig.~\ref{F1} the obtained energy spectrum of some of the
low-lying states and in Table~\ref{Table:E(3)-Spectrum} the values
of the excitation energies of the states, where the energy of the
ground state is set to zero and all energies are normalized to the
energy of the first excited state (with $L^{\pi} = 1^{-}_{1}$).
\begin{figure}[h!]
\begin{center}
\includegraphics[scale=1.0]{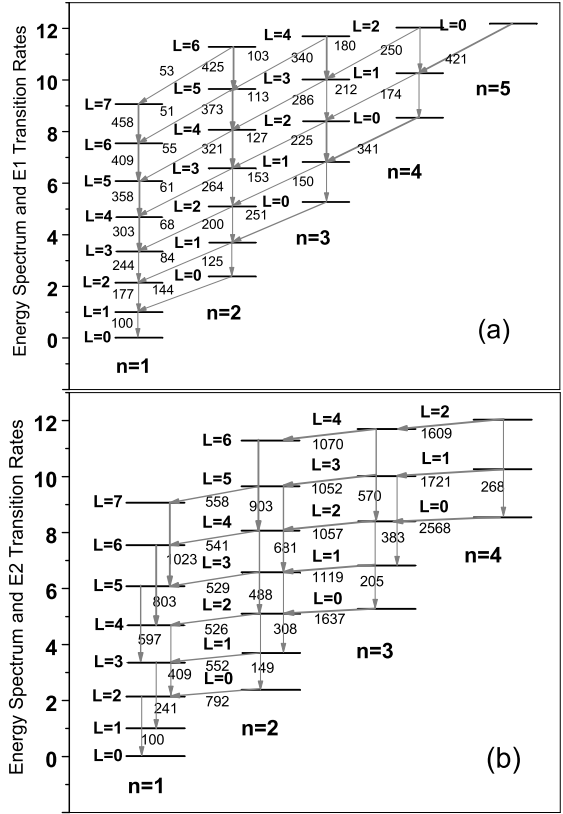}
\vspace*{-5mm} \caption{Calculated energy spectrum of the low-lying
states at the critical point of the phase transition U(3)-O(4) (with
a potential of $r^4$) in the U(4) vibron model and the corresponding
E1 (panel (a)), E2 (panel (b)) transition rates (the numbers marked
close to the arrow are the value of the rate with normalization
$B(\mbox{E1};1^{-}_{1} \rightarrow 0^{+}_{1}) =100$,
$B(\mbox{E2};2^{+}_{1} \rightarrow 0^{+}_{1})=100$, respectively). }
\label{F2}
\end{center}
\end{figure}

\begin{table}[h!]
\begin{center}
\caption{Excitation energies of the low-lying states at the critical
point of the phase transition from U(3) to O(4) in the U(4) vibron
model. } \label{Table:E(3)-Spectrum}
\begin{tabular}{|c|c|c|c|c|c|c|c|c|c|c|}
\hline \hline
{} & \multicolumn{2}{|c|}{$n=1$} &
\multicolumn{2}{|c|}{$n=2$} & \multicolumn{2}{|c|}{$n=3$} &
\multicolumn{2}{|c|}{$n=4$} & \multicolumn{2}{|c|}{$n=5$}
\\ \cline{2-11} {} {} & E(3) & $r^{4}$ & E(3) & $r^{4}$ & E(3)
& $r^{4}$ & E(3) & $r^{4}$ & E(3) & $r^{4}$           \\
\hline $L=0$ & 0.00 & 0.00 & 2.87 & 2.37 & 7.65 & 5.27 & 14.35 &
8.55 & 22.96 & 12.17
\\ \hline
$L=1$ & 1.00 & 1.00 & 4.83 & 3.70 & 10.57 & 6.82 & 18.22 & 10.27 &
27.79 & 14.02
\\ \hline
$L=2$ & 2.26 & 2.13 & 7.06 & 5.09 & 13.76 & 8.41 & 22.37 & 12.02 &
32.90 & 15.89
\\ \hline
$L=3$ & 3.78 & 3.36 & 9.56 & 6.55 & 17.23 & 10.04 & 26.86 & 13.81 &
38.28 & 17.80
\\ \hline
$L=4$ & 5.53 & 4.68 & 12.32 & 8.08 & 20.97 & 11.72 & 31.51 & 15.64 &
43.95 & 19.76
\\ \hline
$L=5$ & 7.51 & 6.09 & 15.34 & 9.67 & 24.97 & 13.47 & 36.48 & 17.52 &
49.89 & 21.75
\\ \hline
$L=6$ & 9.76 & 7.56 & 18.61 & 11.31 & 29.22 & 15.26 & 41.72 & 19.43
& 56.10 & 23.77
\\ \hline
$L=7$ & 12.07 &  9.08    & 22.12 & 13.12 & 33.75 & 16.83&47.22&
21.42& 62.58&25.89
\\ \hline
$L=8$ & 14.90 &  10.71    & 25.88 & 14.75 & 38.52 &19.98& 52.99&
23.41& 69.32&27.99
\\ \hline \hline
\end{tabular}
\end{center}
\end{table}

\begin{figure}[h!]
\begin{center}
\includegraphics[scale=0.9]{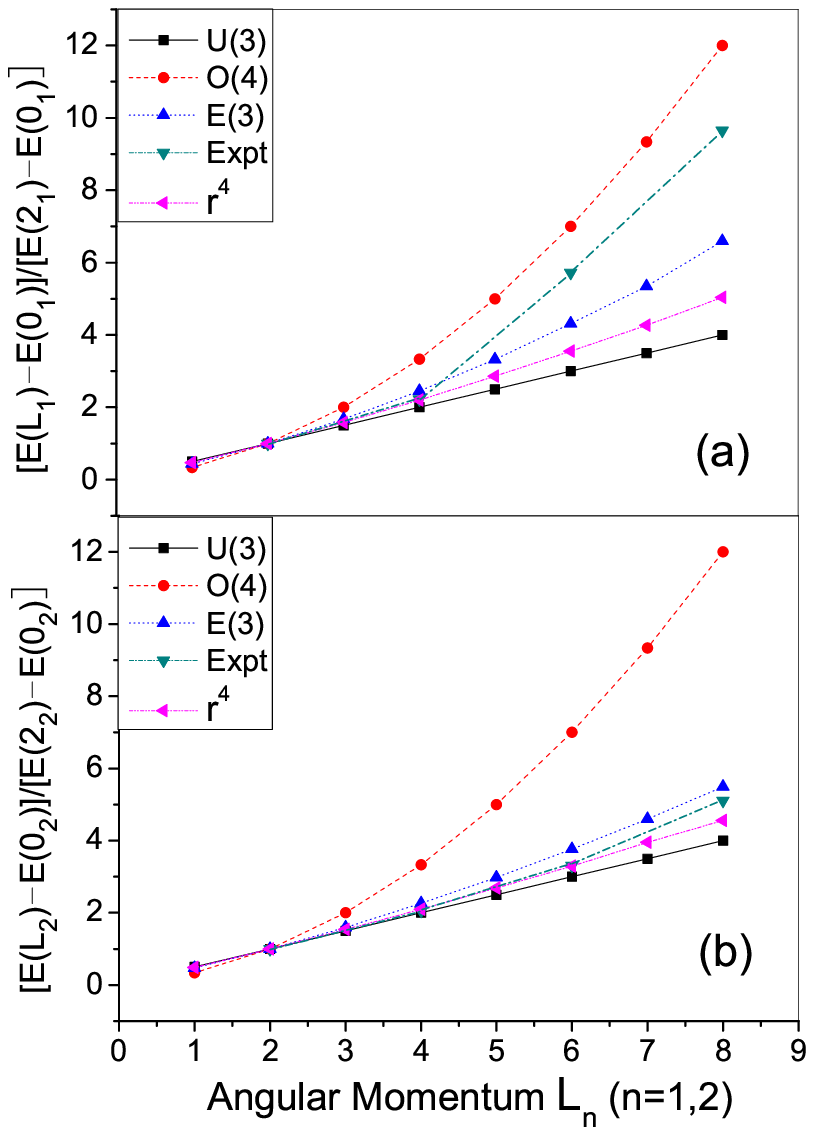}
\vspace*{-5mm} \caption{Calculated ratios of the energies of some
low-lying states $\frac{E(L_{n})-E(0_{n})}{E(2_{n})-E(0_{n})}$ (with
$E(0_{1})=0$ and $n=1$ panel (a), $n=2$ panel (b)) and some of the
experimental data of $^{12}$C+$^{12}$C system. } \label{F3}
\end{center}
\end{figure}

Then we solve the Eq.~(\ref{Schrodinger}) numerically to discuss
more practically the properties of the low-lying states at the
critical point of the phase transition in the U(4) model. The
obtained energy spectrum of the low-lying states is illustrated in
Fig.~\ref{F2} and the concrete values of some states' energies are
listed in Table~\ref{Table:E(3)-Spectrum}. To show the variation
feature of the energy of low-lying states explicitly and make it
easy to compare with experimental data, we display some of the
energy ratios between some states in Fig.~\ref{F3}. Comparing the
result in the E(3) symmetry with that obtained by solving the
Schr\"{o}dinger equation in potential of $r^{4}$ form, one can
notice that, the energy of a state in the E(3) symmetry is higher
than that given by the $r^{4}$ potential for the state with the same
quantum number.

\begin{figure}[h!]
\begin{center}
\hspace*{-5mm}\includegraphics[scale=0.7]{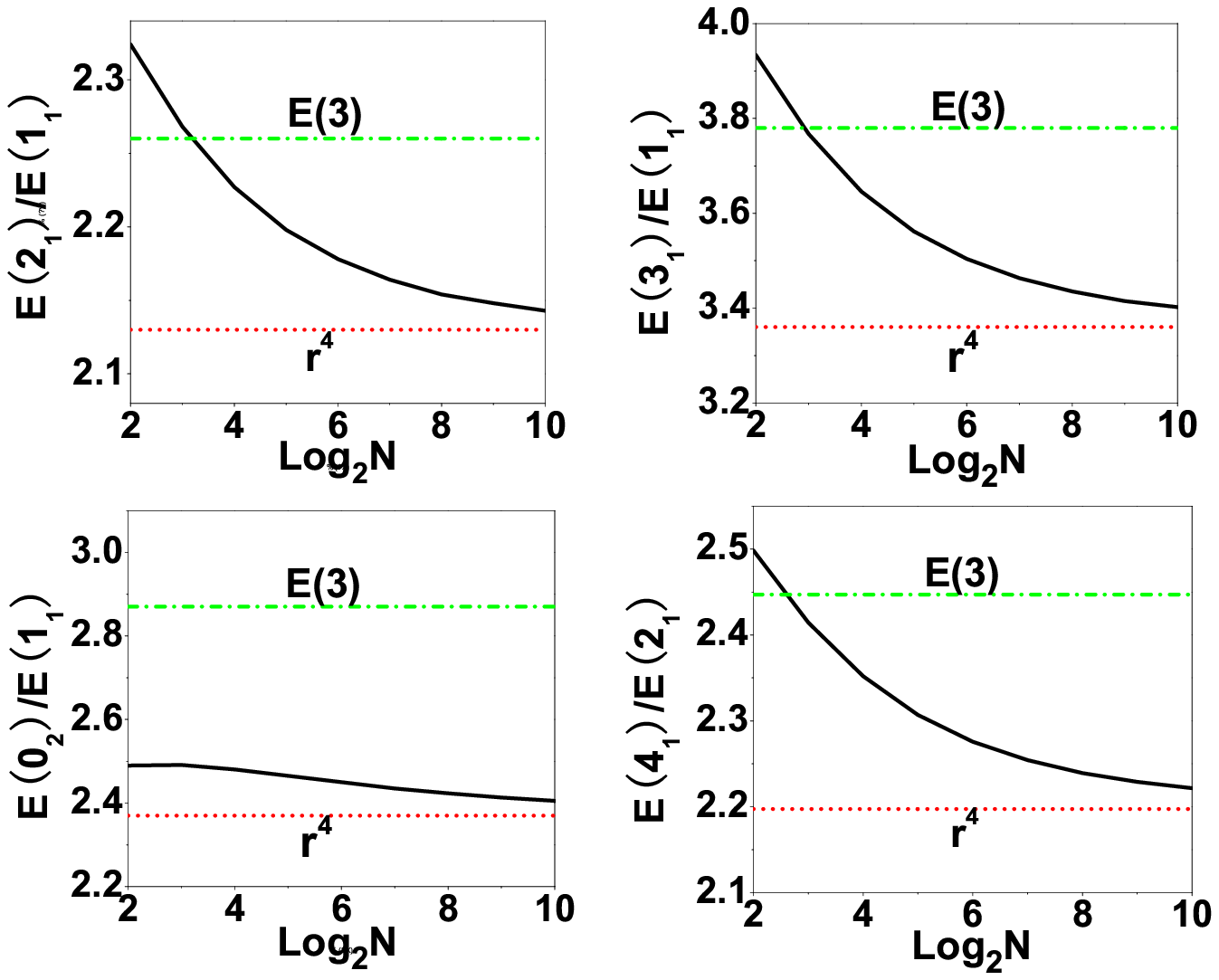}
\vspace*{-10mm}\caption{Calculated variation behavior of the energy
ratios of some low-lying states with respect to the boson number and
those in the E(3) symmetry and in the $r^{4}$ potential. }
\label{F4}
\end{center}
\end{figure}
\begin{figure}[h!]
\begin{center}
\includegraphics[scale=0.9]{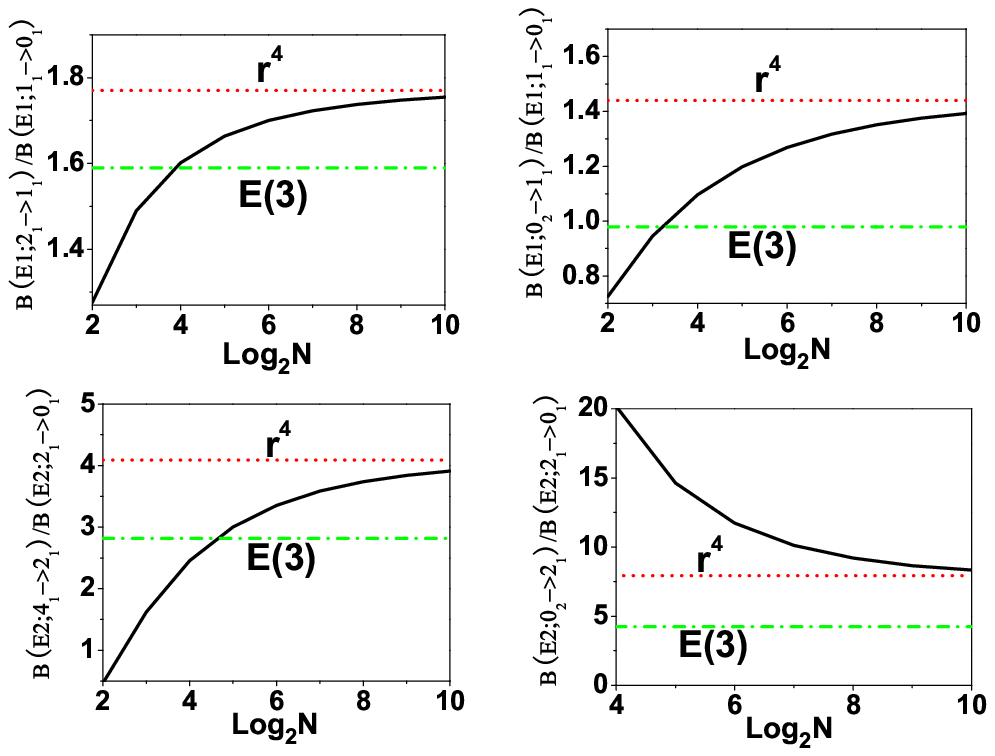}
\vspace*{-5mm}\caption{Calculated variation behavior of the ratios
of some E1, E2 transition rates between low-lying states with
respect to the boson number and those in the E(3) symmetry and in
the $r^{4}$ potential. } \label{F5}
\end{center}
\end{figure}

To investigate the properties of the states at the critical point
more comprehensively, we also solved the engin-equation with
Hamiltonian in Eq.~(\ref{Hamiltonian}) with $\eta = \frac{3}{7}$
directly in the cases of various boson numbers. The obtained results
of the variation feature of the ratios of some low-lying states
against the boson number and the comparison with those in the E(3)
symmetry and $r^{4}$ potential are illustrated in Fig.~\ref{F4}.
From Fig.~\ref{F4}, one can learn that, for the yrast states, the
results given in the E(3) symmetry is quite close to that in the
case of fewer bosons (for instance, around ten), but that determined
by the $r^{4}$ potential is the asymptotic limit of the one in large
boson number limit. It indicates that the one with the $r^{4}$
potential describes the behavior at the classical limit (or infinite
boson number limit) more appropriately, and the E(3) symmetry
over-predicts the energy levels. Such a feature is consistent with
those of the states around the critical point of the phase
transition from U(5) to O(6) in the U(6) model of
IBM~\cite{Arias2003,Ramos2005}.

With transition operators
\begin{equation}
\hat{T}(\mbox{E1}) = - \vec{P} \cdot \vec{E}_0 = e_{1,eff}
r\cos{\theta} \, ,
\end{equation}
\begin{equation}
\hat{T}(\mbox{E2}) = -\frac{1}{6}\sum_{ij}Q_{ij}\Big( \frac{\partial
E_j}{\partial x_i } \Big)_{0} = e_{2,eff} r^2 ( 3\cos^2{\theta} - 1)
\, ,
\end{equation}
where $Q_{ij} = q \langle 3 x_i x_j - r^2 \delta_{ij} \rangle $, we
can determine the E1 and E2 transition rates $B(\mbox{Ek}, L_{i}
\rightarrow L_{f} ) = \frac{1}{2L_{i} + 1} \vert \langle n_{f} \,
L_{f} \Vert \hat{T}(\mbox{Ek}) \Vert n_{i} \, L_{i} \rangle \vert
^2$.
The obtained results in case of the E(3) symmetry (i.e., with the
potential of infinite square well) and those in potential of $r^{4}$
are displayed in Fig.~\ref{F1}, Fig.~\ref{F2}, respectively. We have
also calculated the E1 and E2 transition rates with operator
$\hat{T}(E1)_{q} =e_{1,eff} (s^{\dag} \tilde{p} + p^{\dag}
s)^{1}{q}$, $\hat{T}(E2)_{\mu} =e_{2,eff} (p^{\dag}
\tilde{p})^{2}{\mu}$, respectively, for the systems with different
boson numbers. The calculated results of changing behavior of the
ratios of some transition rates with respect to the boson number and
the comparison with those in E(3) symmetry and $r^{4}$ potential are
manifested in Fig.~\ref{F5}. Looking through Figs.~\ref{F1},
\ref{F2} and \ref{F5}, one can notice that the E(3) symmetry
under-predicts the E1 and E2 transitions rates between the low-lying
states at the critical point with the $r^{4}$ potential in the U(4)
model. Moreover, the E(3) symmetry can only describe the ratios of
some transition rates of the system with boson number $N$ around 10
to 20 with an exception of
$B(E2:0_2\rightarrow2_1)\over{B(E2:2_1\rightarrow 0_1)}$, which
deviates from both the result in E(3) symmetry and that in the
$r^{4}$ potential in small $N$ case but in large-$N$ limit obviously
approaches to the one in the $r^{4}$ potential. Therefore, the
$r^{4}$ potential depicts excellently all the transition rate ratios
as those energy ratios of the system in classical limit. The
analyses of energy levels together and transition rates indicate
that the results in the E(3) symmetry shows a nature finite $N$
correction on those in the $r^{4}$ potential for the low-lying
spectrum since the bottom of the potential in E(3) symmetry is
flatter than that in the $r^{4}$ potential. As for the high-lying
spectrum, this conclusion may be changed, which needs to be further
tested.

\begin{table}[h!]
\caption{Experimental data of the energy ratios of some low-lying
resonant states of ${^{12}}$C+${^{12}}$C and the corresponding
values in the $r^{4}$-potential, E(3) symmetry, U(3) symmetry and
O(4) symmetry (the experiment data are taken from
Ref.~\cite{Hess2003}).} \label{Table:CP-C} \vspace*{-6pt}
\begin{center}\begin{tabular}{ccccccc}\hline\hline
~~~~~~~~~~~~~~& ~~Expt.~~ & ~~ E(3)~~ & ~~ $r^{4}$ ~~ & ~~ U(3) ~~ &
~~ O(4) ~~ \\\hline
$E_{4_1}/E_{2_1}$& 2.22 & 2.45 & 2.20 & 2 & $\frac{10}{3}$\\
$E_{6_1}/E_{2_1}$& 5.72 & 4.31 & 3.55 & 3 & 7\\
$E_{0_2}/E_{2_1}$& 0.31 & 1.27 & 1.05 &1&$\frac{2N}{3}$\\
$E_{2_2}/E_{2_1}$& 2.50 & 3.12 & 2.39 & 2 &$\frac{2N}{3}+1$\\
$E_{4_2}/E_{2_1}$& 4.48 & 5.45 & 3.79 & 3 &$\frac{2N}{3}+\frac{10}{3}$\\
$E_{6_2}/E_{2_1}$& 7.55 & 8.22 & 5.31 & 4 &$\frac{2N}{3}+7$\\
$E_{0_3}/E_{2_1}$& 1.86 & 3.38 & 2.47 & 2 & $\frac{4N-4}{3}$  \\
$E_{4_1}/E_{0_2}$& 7.16 & 1.93 & 1.97 & 2 &$\frac{5}{N}$\\
$E_{6_1}/E_{0_2}$&18.45 & 3.39 & 3.19 & 3 &$\frac{21}{2N}$\\
$E_{4_1}/E_{0_3}$& 1.19 & 0.72 & 0.89 & 1 & $\frac{5}{2N-2}$ \\
$E_{6_1}/E_{0_3}$& 3.08 & 1.27 & 1.43 & $1.5$ & $\frac{21}{4N-4}$ \\
\hline\hline
\end{tabular}
\end{center}
\end{table}

It has been shown that the U(4) model can successfully describe
nuclear molecules~\cite{Iachello19846,Hess2003}. A nuclear vibron
model for nuclear molecules consisting of two clusters holds
generally a dynamical symmetry $G_{C1}\otimes G_{C2}\otimes
U_{R}(4)$, where the internal structure of the $i$th cluster is
described by $G_{Ci}$ which may be, for example, the U(6) IBM or
SU(3) shell model, and the relative motion between the clusters is
described by the U$_{R}$(4) vibron model. In some cases, only the
U(4) vibron model itself is sufficient to describe the rotational
and vibrational excitations in nuclear molecules, where the internal
structure of each cluster does not play an essential role in the
low-lying levels, such as, the narrow resonances in the
${^{12}}$C+${^{12}}$C system~\cite{Erb1981,Cseh1985}. In the early
time, the O(4) limit of U(4) vibron model was proposed to describe
the resonant energy of ${^{12}}$C+${^{12}}$C system~\cite{Erb1981},
while the analysis in Ref.~\cite{Cseh1985} indicates that the U(3)
limit may be more preferred when fitting the energies of the
low-lying resonant states, and Ref.~\cite{Hess2003} shows that the
experimental data are in fact between those with U(3) and O(4)
symmetry, respectively. Recently some of us proposed that the
${^{12}}$C+${^{12}}$C system may be described by the critical
symmetry, E(3), in the U(3)-O(4) phase transition~\cite{Liu2008}.
Since the potential at the critical point of the U(3)-O(4) phase
transition is in fact in the form of $r^{4}$ but not that in the
E(3) symmetry, to show the practical possibility of such a system
being that at the critical point of the phase transition in the U(4)
model, we should re-analyze that more cautiously. Then we display
the experimental data of some ratios of the low-lying resonant
energies of the ${^{12}}$C+${^{12}}$C system
as well as the corresponding results with U(3), O(4), E(3) symmetry
and $r^{4}$-potential in Table~\ref{Table:CP-C} and Fig.~\ref{F3}.
One can find from Table~\ref{Table:CP-C} and Fig.~\ref{F3} that the
experimental data agree globally better with the results in
$r^{4}$-potential than with those in the U(3), O(4) and E(3)
symmetries.
It indicates that, concerning energies of the low-lying resonant
states, the ${^{12}}$C+${^{12}}$C system is an empirical evidence of
the states at the critical point of the U(3)-O(4) phase transition
in the U(4) model, and the E(3) symmetry can describe that
approximately.


In summary, we have calculated the energy spectra and the transition
rates of not only E1 but also E2 transitions of the low-lying states
at the critical point of the phase transition from U(3) to O(4) in
the U(4) vibron model in coordinate space with both the E(3)
symmetry (or with potential in the form of infinite well) and the
$r^{4}$ potential as well as in boson space with finite $N$. Our
calculation shows that the large-$N$ limit of the U(4) vibron model
at the critical point of phase transition can be represented by a
$r^{4}$ potential model excellently, but not close to those of an
infinite well in the $E(3)$ symmetry. Generally, The E(3) symmetry
over-predicts the energy levels and under-predicts the E1 and E2
transition rates of the states at the critical point of the phase
transition in the U(4) model in large-$N$ limit, but predicts a
critical-point spectrum that is qualitatively similar to the U(4)
vibron model for small values of $N$ in the most cases. The
calculated results also express that the ratios of quantities at the
critical point quickly approach to a constant with the increasing of
boson number $N$ when $N>30$, which further confirms the $N$ scaling
behavior of quantities at the critical point shown in
Ref.~\cite{Liu2008}, where the results indicate that the ratios of
both energies and E1 transition at the critical point are
approximately invariant as $N$ varies. Comparing the theoretical
results with the experimental data of some systems, we find that the
${^{12}}$C+${^{12}}$C system is an empirical evidence of the state
at the critical point of the U(3)-O(4) phase transition in the U(4)
model and the E(3) symmetry can describe that roughly when
concerning the energies of the low-lying resonant states. To conform
it much more solidly, one needs the data of electromagnetic
transition rates.

\bigskip

This work was supported by the National Natural Science Foundation
of China under the Grant Nos. 10425521 and 10675007, the National
Fund for Fostering Talents of Basic Science (NFFTBS) with contract
No. J0630311, the Major State Basic Research Development Program
under Contract No. G2007CB815000, the Key Grant Project of Chinese
Ministry of Education under contact No. 305001.


\newpage

\begin{thebibliography}{100}

\bibitem{GK1980}J. N. Ginocchio, and M. W. Kirson,
 Phys. Rev. Lett {\bf 44}, 1744 (1980).

\bibitem{DSI1980}A. E. L. Dieperink, O. Scholten, and F. Iachello,
            Phys. Rev. Lett. {\bf 44}, 1747 (1980).

\bibitem{FGD1981}D. H. Feng, R. Gilmore, and S. R. Deans, Phys. Rev.
         C {\bf 23}, 1254 (1981).

\bibitem{IC1981}P. Van Isacker, and J. Q. Chen, Phys. Rev. C {\bf 24},
         684 (1981).

\bibitem{WC1983}D. D. Warner, and R. F. Casten, Phys. Rev. C {\bf 28},
         1798 (1983).

\bibitem{IachelloBook87}F. Iachello, and A. Arima, {\it The Interacting Boson
Model} (Cambridge University, Cambridge, England, 1987).

\bibitem{Leviatan} A. Leviatan, Phys. Rev. Lett. {\bf 77}, 818 (1996);
A. Leviatan, and P. Van Isacker, Phys. Rev. Lett. {\bf 89}, 222501
(2002); A. Leviatan, Phys. Rev. Lett. {\bf 98}, 242502 (2007).

\bibitem{Warner02}
D. Warner, Nature {\bf 420}, 614 (2002).

\bibitem{Jolie2002}J. Jolie, P. Cejnar, R. F. Casten, S. Heinze, A. Linnemann,
         and V. Werner, Phys. Rev. Lett. {\bf 89}, 182502 (2002).

\bibitem{LG03}
A. Leviatan and J. N. Ginocchio, Phys. Rev. Lett. {\bf 90}, 212501
(2003).

\bibitem{Regan03} P. H. Regan {\it et al.}, Phys. Rev. Lett. {\bf 90}, 152502 (2003).

\bibitem{Panetc035} F. Pan, J. P. Draayer, and Y. A. Luo, Phys. Lett. B {\bf 576}, 297
(2003); F. Pan, Y. Zhang and J. P. Draayer. J. Phys. G {\bf 31},
1039 (2005).

\bibitem{Iachello2004}F. Iachello and N. V. Zamfir,
         Phys. Rev. Lett. {\bf 92}, 212501 (2004).

\bibitem{Rowe2004}D. J. Rowe, Phys. Rev. Lett. {\bf 93}, 122502
         (2004); D. J. Rowe, P. S. Turner and G. Rosensteel,
         Phys. Rev. Lett. {\bf 93}, 232502 (2004).

\bibitem{Heenen05}
S. Hwiok, P.-H. Heenen, and W. Nazarewicz, Nature {\bf 433}, 705
(2005).

\bibitem{Cejnar2005}P. Cejnar, S. Heinze, and J. Dobe\v{s},
         Phys. Rev. C {\bf 71}, 011304(R) (2005).

\bibitem{Dusuel2005}S. Dusuel, J. Vidal, J. M. Arias, J. Dukelsky,
         and J. E. Garc\'{i}a-Ramos, Phys. Rev. C {\bf 72}, 064332 (2005).

\bibitem{Casten067}
R.F. Casten, Nature Phys. {\bf 2}, 811 (2006); R.F Casten, and E.A.
McCutchan, J. Phys. G {\bf 34}, R285 (2007).
\bibitem{Ring07}
J. Meng, W. Zhang, S. Q. Zhang, H. Toki, and L. S. Geng, Eur. Phys.
J. A {\bf 25}, 23 (2005); T. Nik\v{s}i\'{c}, D. Vretenar, G.A.
Lalazissis, and P. Ring, Phys. Rev. Lett. {\bf 99}, 092502 (2007).

\bibitem{Liu2006}Y. X. Liu, L. Z. Mu, and H. Wei, Phys. Lett. B {\bf 633},
          49 (2006); Y. Zhao, Y. Liu, L.Z. Mu, and Y.X. Liu, Int. J. Mod. Phys.
          E {\bf 15}, 1711 (2006).

\bibitem{Arias2007}J. M. Arias, J. Dukelsky, J. E.
Garc\'{\i}a-Ramos, and J. Vidal, Phys. Rev. C {\bf 75}, 014301
(2007).

\bibitem{Liu2007} Y. Zhang, Z. F. Hou, and Y. X. Liu, Phys. Rev. C {\bf 76},
011305(R) (2007).

\bibitem{Bonatsos08}
D. Bonatsos, E. A. McCutchan, R. F. Casten, and R. J. Casperson,
Phys. Rev. Lett. {\bf 100}, 142501 (2008); D. Bonatsos, E. A.
McCutchan, and R. F. Casten, Phys. Rev. Lett. {\bf 101}, 022501
(2008).


\bibitem{IachelloBook95}F. Iachello and R. D. Levine, {\it Algebraic Theory of
          Molecules} (Oxford University, Oxford, England, 1995)

\bibitem{Kuyucak1999}S. Kuyucak, Chem. Phys. Lett. {\bf 301}, 435 (1999)


\bibitem{Iachello2005}F. P\'{e}rez-Bernal, L. F. Santos, P. H. Vaccaro,
          and F. Iachello, Chem. Phys. Lett. {\bf 414}, 398 (2005)

\bibitem{Hess2006} H. Y\'{e}pez-Mart\'{i}nez, J. Cseh, and P.
          O. Hess, Phys. Rev. C {\bf 74}, 024319 (2006).

\bibitem{Iachello1982}F. Iachello, and R. D. Levine, J. Chem. Phys. {\bf 77},
3046 (1982)

\bibitem{Kuyucak958}S. Kuyucak and M. K. Roberts, Chem. Phys. Lett.
{\bf 238}, 371 (1995); S. Kuyucak and M. K. Roberts, Phys. Rev. A
{\bf 57}, 3381 (1998).

\bibitem{Iachello1981}F. Iachello, Phys. Rev. C {\bf 23}, 2778( 1981).

\bibitem{Erb1981}K. A. Erb, and D. A. Bromley, Phys. Rev. C {\bf 23}, 2781
(1981).

\bibitem{Cseh1985}J. Cseh, Phys. Rev. C {\bf 31}, 692 (1985).

\bibitem{Cseh1993}J. Cseh, G. L\'{e}vai, and W. Scheid, Phys. Rev. C
{\bf 48}, 1724 (1993).

\bibitem{Iachello1991}F. Iachello, N. C. Mukhopadhyay, and L. Zhang, Phys.
Rev. D {\bf 44}, 898 (1991).

\bibitem{Pan2006}F. Pan, Y. Zhang, and J. P. Draayer, Eur. Phys. J. A {\bf 28},
313 (2006).

\bibitem{Dieperink1982}O. S. Van Roosmalen, and A. E. L. Dieperink, Ann. Phys.
(N.Y) {\bf 139}, 198 (1982).


\bibitem{Liu2008} Y. Zhang, Z. F. Hou, H. Chen, H. Q. Wei, and Y. X. Liu,
Phys. Rev. C {\bf 78}, 024314 (2008).

\bibitem{Iachello2000}F. Iachello, Phys. Rev. Lett. {\bf 85}, 3580 (2000).

\bibitem{Iachello2001}
F. Iachello, Phys. Rev. Lett. {\bf 87}, 052502 (2001).

\bibitem{Arias2003} J. M. Arias, C. E. Alonso, A. Vitturi, J. E. Garc\"{i}a-Ramos,
J. Dukelsky, and A. Frank, Phys. Rev. C {\bf 68}, 041302(R) (2003).

\bibitem{Bonatsos2004} D. Bonatsos, D. Lenis, N. Minkov, P.P. Raychev,
and P. A. Terziev, Phys. Rev. C {\bf 69}, 044316 (2004).

\bibitem{Ramos2005} J. E. Garc\'{i}a-Ramos, J. Dukelsky, and J. M. Arias,
Phys. Rev. C {\bf 72}, 037301 (2005).

\bibitem{McCutchan2006} E. A. McCutchan, D. Bonatsos, and N. V.
Zamfir, Phys. Rev. C {\bf 74}, 034306 (2006).

\bibitem{Arias2008} J. E. Garc\"{i}a-Ramos, and J. M. Arias, Phys. Rev. C
{\bf 77}, 054307 (2008).

\bibitem{Casten00}
  R. F. Casten, and N. V. Zamfir, Phys. Rev. Lett. {\bf 85}, 3584 (2000);
  A. Frank, C. E. Alonso, and J. M. Arias, Phys. Rev. C {\bf 65}, 014301 (2002);
  D. L. Zhang, and Y. X. Liu, Phys. Rev. C {\bf 65}, 057301 (2002);
  R. M. Clark {\it et al.}, Phys. Rev. C {\bf 69}, 064322 (2004); H. von Garrel {\it et al.},
  Phys. Rev. C {\bf 73}, 054315 (2006).
\bibitem{Casten01}
  R. F. Casten, and N. V. Zamfir, Phys. Rev. Lett. {\bf 87}, 052503 (2001);
  R. Krucken {\it et al.}, Phys. Rev. Lett. {\bf 88}, 232501 (2002);
  C. Hutter {\it et al.}, Phys. Rev. C {\bf 67}, 054315 (2003);
  D. L. Zhang, and Y. X. Liu, Chin. Phys. Lett. {\bf 20}, 1028 (2003);
  D. Tonev {\it et al.}, Phys. Rev. C {\bf 69}, 034334 (2004);
  A. Dewald, et al., Eur. Phys. J. A {\bf 20}, 173 (2004);
  O. M\"{o}ller {\it et al.}, Phys. Rev. C {\bf 74}, 024313 (2006);
  A. F. Mertz {\it et al.}, Phys. Rev. C {\bf 77}, 014307 (2008).

\bibitem{Iachello19846}F. Iachello, Nucl. Phys. A {\bf 421}, 97c
(1984); H. J. Daley and F. Iachello, Ann. Phys. (N.Y.), {\bf 167},
73 (1986).

\bibitem{Hess2003}H. Y\'epez-Mart\'{i}nez, P. O. Hess, and S. Misicu,
Phys. Rev. {\bf C 68}, 014314 (2003).

\end{thebibliography}
\end{document}